\documentclass{PoS}
\newcommand{\dF}{{^{^*}\!\!F}}

\title{Multi-messenger signals from short gamma ray bursts}

\ShortTitle{Short GRBs}

\author{\speaker{Agnieszka Janiuk}\thanks{Special thanks to Irek Janiuk and my family for all support}\\
  Center for Theoretical Physics\\
  Polish Academy of Sciences\\
  Al. Lotnikow 32/46, 02-668 Warsaw, Poland\\
        E-mail: \email{agnes@cft.edu.pl}}
\author{Kostas Sapountzis\\
  Center for Theoretical Physics\\
  Polish Academy of Sciences\\
  Al. Lotnikow 32/46, 02-668 Warsaw, Poland\\
}
\author{Bestin James\\
  Center for Theoretical Physics\\
  Polish Academy of Sciences\\
  Al. Lotnikow 32/46, 02-668 Warsaw, Poland\\
}
\author{Martin Kolos\\
  Institute of Physics\\
  Faculty of Philosophy and Science, Silesian University \\
  Bezrucovo nam. 13, CZ-74601, Opava, Czech Republic\\
}  

\abstract{We present the results of simulations done with the code HARM-COOL
  developed in the CTP PAS Warsaw research group over the years 2017-2019. It is based in the original GR MHD scheme proposed by Gammie et al.
  (2003) for the simulation of Active Galactic Nucleus, but now it has been suited for the engine of a short Gamma Ray Burst event.
  We compute time-dependent evolution of a black hole accretion disk, in two-dimensional, axisymmetric scheme.  The code includes
  neutrino cooling and accounts for nuclear structure of dense, degenerate matter. Free protons, neutrons, and electron-positron
  pairs form a neutron-rich, magnetically driven outflow that provides site for subsequent r-process nucleosynthesis.
  Here the heavy elements up to the Uranium and Gold are synthesized and may contribute to the chemical enrichment of the circum-burst medium. Their radio-active decay will give signal in lower energies in a timescale of weeks-months after the GRB prompt phase.
  In addition, the magnetic fields are responsible for the launching of ultra-relativistic jets along the
  rotation axis of the central black hole, according to the well-known Blandford-Znajek mechanism. These jets are sites of variable high energy emission in gamma rays. We find that the magnetic field and the black hole spin account for the observed variability timescales and jet energetics.
 }

\FullConference{Multifrequency Behaviour of High Energy Cosmic Sources - XIII - MULTIF2019\\
		3-8 June 2019\\
		Palermo, Italy}

\begin{document}

\section{Introduction}

\subsection{Gamma Ray Bursts}

Gamma Ray bursts are rapid flashes of high energy radiation, that is typically
peaking in the soft gamma-ray and hard X-ray band. The events occur at an average rate of one event per day, and originate from cosmological distances.
Observed energies of photons are on the order of Mega-electronovolt, while the time variability
in the millisecond scale suggests that the emission is produced in a very compact region.

There is the definition of the so called $T_{90}$, which gives duration of GRBs is that of which 5-95\% of the total fluence is
detected. Hence, the exact duration of the event is affected by the sensitivity of detector at a
given energy band.
Nevertheless, the classification suggested by \cite{kouval1993}
provided two basic categories of GRBs, with $T_{90}<2$ s
in case of short events, and $T_{90}>2$ s in the long ones.
These two classes correspond to two different progenitor types, namely the long GRBs
are produced in the cataclysmic explosions of massive stars, and short GRBs are related to the
mergers of binary compact objects.

The spectral energy distributions of GRBs follow the broken power-law \cite{band1993} function,
where the peak energy, $E_{p}$, is about several hundred keV.
The spectra of prompt emission rapidly vary with time,
and the pulses observed in softer bands are lagged by these in harder bands \cite{zhang_review} (it should be noted however that a number of GRBs, both long and short, are found with negligible spectral lag, see \cite{bernardini2015}).

At other wavelengths, the afterglow emission is detected with lower energies, while
also some Optical emission related to the prompt GRB phase was reported in a few cases.
No detection so far was reported as for the high energy neutrinos associated with GRBs, and only upper limits are available (see \cite{guepin2017} and references therein).
Finally, the Optical and near Infrared emission, with the lightcurves profile
explained by the radioactive decay of
unstable heavy isotopes, was reported twice. First tentative associaction of GRB 130603B
\cite{Tanvir2013} was attributed to the 'kilonova' emission, and recently the observations
of GW 170817/GRB 170817A \cite{abbott2017} confirmed these findings for the short GRB.
These heavy isotopes are presumably synthesized in the ejecta released during binary neutron star merger
that is the short GRB progenitor \cite{li_paczynski98}.

\subsection{Central engines and progenitors}

Time variability, pulse profiles, and total energetics of the GRBs indicate that they are emitted by a
collimated relativistic outflow of plasma, that is pushing through the interstellar medium.
These outflows, called jets, are powered by the central engine.
Similarly to the other types of sources with astrophysical jets, the plausible and the popular model for
such an engine is black hole (or a magnetar), surrounded by accreting material in a form of torus, that is mediated by
magnetic fields and extracts the rotational energy of the central black hole \cite{bz77}.
In addition to that, the source of power in case of GRB jets might be the anihillation of neutrino-antineutrino pairs, produced inside the torus. Here, the process might have a possibly smaller efficiency \cite{janiuk2017}.

Both long and short GRBs to some extent share the properties of their central engines.
In long GRBs, the collapse of a rotating massive star into a black hole is forming the engine. Predicted is the beamed explosion, accompanied by a supernova-like ejection. This theoretical prediction is supported by the fact that the long GRB lightcurves show a power-law decay profile, which is characteristic for the relativistic blast wave explosions. Also, the hosts of long GRBs are located in star forming regions,
which was indicated by their high gas densities and metallicity \cite{holland2001, chevalier2004}. The emission lines characteristic for the underlying
supernova explosion were detected in the spectra of many events, starting from \cite{stanek2003}.

For the short GRBs, the progenitor hosts are more difficult to study,
because of statistically fewer events which have adequately good data for
the afterglow detection. Therefore, the leading candidates for progenitors,
namely the
neutron star (NS-NS) mergers, have been studied mainly by numerical
simulations, e.g., \cite{korobkin2012, rezzolla2011, paschalidis},
and by population synthesis analyses \cite{mapelli2018, oslowski2011}.
An alternative explanation for the progenitor candidates can be the accretion induced collapse of neutron star, or the giant flare of soft gamma-ray repeaters \cite{nakar2006}.
Only the recent detection of the short GRB associated with the gravitational wave, whose for is consistent with that released during the inspiral and merger of two neutron stars, has proven the NS-NS progenitor scenario. The Fermi-GBM observation
was reported by \cite{goldstein2017} at 1.7 seconds after the LIGO signal, and
indicates that the system must have gone also through the transient phase of the
hyper-massive neutron star (HMNS) \cite{baiotti_rezzolla2017}.

\section{Methods of study of the central engine}

The models of GRB central engine describe the accretion flow onto a black hole.
Any physical model must satisfy the basic equations of hydrodynamics: (i)
the continuity equation, (ii) the energy equation,
and (iii) the conservation of momentum
(accounts for the radial transport, and rotation - hence the angular momentum is conserved).
This has to be supplemented with the equation of state.
In the simplest case, the equation of state is that of an ideal gas.
The model must also describe the transport of angular momentum, if the
flow rotates at Keplerian velocity and matter is to be accreted.
In simplest case, one can use the $\alpha$-disk model, where the stress
scales with the gas pressure \cite{ss73}.
%(Shakura \& Sunyaev 1973).
This prescription successfully mimics the angular momentum 
transport by magneto-hydrodynamical
(MHD) turbulences and also prescribes the viscous dissipation of energy.
The $\alpha$-disk models were applied to the GRB central engine problem
by e.g. \cite{popham1999, dimatteo2002, janiuk2004, lei2009}.

\subsection{HARM code}

Our current simulations were made with 2D version of the code {\it HARM} 
The acronym stands for the High Accuracy Relativistic Magnetohydrodynamics
\cite{gammie2003}.
%(Gammie et al. 2003). 
The code provides solver for continuity and energy-momentum conservation equations in GR:
\[ \nabla_{\mu}(\rho u^{\mu}) = 0 ~~~~~ \nabla_{\mu}T^{\mu\nu} = 0\]
%\[ {T^\mu}_{\nu;\mu} = 0 \]

The Energy tensor contains in general electromagnetic and gas parts:
\[ T^{\mu\nu} = T^{\mu\nu}_{\rm gas} + T^{\mu\nu}_{EM} \]
\[ T^{\mu\nu}_{gas} = \rho h u^{\mu} u^{\nu} + pg^{\mu\nu} =(\rho + u + p) u^{\mu} u^{\nu} + pg^{\mu\nu} \]
\[ T^{\mu\nu}_{EM} = b^{2} u^{\mu} u^{\nu} + \frac{1}{2}b^2 g^{\mu\nu} - b^{\mu} b^{\nu} \]
with $b^{\mu} = u_{\nu}\dF^{\mu\nu}$,
where $u^\mu$ is four-velocity of gas, $u$ is internal energy density, and 
$b^\mu = \frac{1}{2} \epsilon^{\mu\nu\rho\sigma}u_\nu F_{\rho\sigma}$
and $F$ is the electromagnetic tensor.
In the force-free approximation, we have $E_{\nu}=u_{\mu}F^{\mu\nu}=0$.
EOS in simplest case is that of ideal gas, and in the code it is using the adiabatic index $\gamma$:
\[ p = K\rho^\gamma = (\gamma-1) u \]

\section{Structure of the GRB central engine}

The central engine in GRB is composed of the rotation,
stellar-mass black hole, accretion disk, and magnetized Poynting-dominated jet.
To model this structure, we adopt an initial model of a rotationally-supported torus
around a black hole in the Kerr metric, and
an initial configuration of the magnetic field. The field provides mechanism for
transport of the angular momentum, and therefore it allows the matter to accrete through the black hole horizon, even if the initial configuration was based on some
equilibrium solution.
The code transforms coordinates from the standard Boyer-Lindquist, in which only the initial problem is specified,
to the Kerr-Schild coordinates, which are non-singular on the horizon, and are defined as $\left(t,r,\theta,\phi\right)$, with the line element given by:
\begin{eqnarray*}
  ds^{2} = - \left({1 - {2r \over R^{2}}} \right) dt^{2} + \left({4r \over R^{2}} \right) dr dt + \left({1 + {2r \over R^{2}}}\right) dr^{2} + R^{2} d\theta^{2}
\\
  + \sin^{2}\theta \left[ R^{2} + a^{2} \left({1 + {2r \over R^{2}}} \right) \sin^{2}\theta \right] d\phi^{2}
        - \left( { 4ar \sin^{2}\theta \over R^{2}} \right) d\phi^{2}
        - 2a \left( {1 + {2r \over R^{2}}} \right) \sin^{2}\theta dr d\phi   
\end{eqnarray*}
where $R=r^{2}+a^{2}\cos^{2}\theta$, and $c=G=M=1$ is assumed.
In addition, the MKS (Modified Kerr-Schild) coordinates $\left(t,x^{(1)},x^{(2)},\phi\right)$ are obtained
with the transformation:
%\cite{Noble2006}:
\begin{eqnarray*}
r= R_0 + \exp\left[{x^{(1)}}\right]
\\
\theta = \frac{\pi}{2} \left(1 + x^{(2)}\right) + \frac{1 - h_{\rm slope}}{2} \sin\left[\pi\left(1 +x^{(2)}\right)\right]
\end{eqnarray*}
where $R_0$ is the innermost radius of the grid, and $h_{\rm slope}$ is a
parameter that determines the concentration of points at the mid-plane.
In our models we use $h=0.3$, and $R_0=0.$

\subsection{Initial conditions}

%equations
%torus
%magnetic fields
%some illustrations
As the initial condition for the simulation, we consider the
toroidal perfect fluid configuration around rotating Kerr BH.
The model is called sometimes the 'Polish doughnut' and has been proposed in a number of woks
\cite{FishMon1976, Abr1978}.
Here, the barotropic equation of state $p=p(\rho)$ is assumed, and the matter is in orbital motion only, so $u^{\theta}=0$ and $u^r=0$. The Euler equation
can be written as an equation for the barotropic pressure $p(\varrho)$ as follows:
\begin{equation}
\frac{\partial_{\mu}p}{\varrho+p}=-{\partial_{\mu }W}+\frac{\Omega \partial_{\mu} l}{1-\Omega \ell},  \quad
W\equiv -\ln\left( -g_{tt} -2\Omega g_{t\phi}  -g_{\phi\phi} \Omega^2 \right) + l_* \Omega 
\label{diskEQ}
\end{equation}
where $l_* = l/(1-\Omega l)$ is constant through the accretion torus, $\Omega=u^{\phi}/u^{t}$ is the fluid relativistic angular frequency related to distant observers, while $W(r;\ell,a)$ is the Paczy{\'n}ski-Wiita potential. The fluid equilibrium is regulated by the balance of the gravitational and pressure terms versus centrifugal factors arising due to the fluid  rotation and the curvature effects of the Kerr background. 
The value of specific angular momentum, $l_*$, is uniquely determined in FM torus by the radius of the pressure maximum. In our computations, we adopt $r_{\rm max}=(7.5 - 25) r_{\rm g}$, which corresponds to $l_*=(3.85-5.55)$. (Here, $r_{\rm g}=GM/c^{2}$ is the gravitational radius of the black hole.)
Other initial configurations of matter distribution are also possible, e.g.,
\cite{chakrabarti85}.

For the magnetic fields,  we introduce initially
the poloidal field configuration, given by the third component of the vector potential, $A_{\phi}(r,\theta)$.
The time evolution and relaxation of magnetized matter into more realistic configuration, creating also the toroidal component of the magnetic field
with time, is the result of the subsequent
GRMHD simulation. 
  
Initial magnetic field can be adopted as that
of an electric wire (see \cite{paschalidis}):
%(cf. Paschalidis et al. 2016), i.e.
\begin{eqnarray}
  A_{\phi}({r,\theta})= A_0 \frac{\left(2-k^2\right) K\left(k^2\right)-2 E\left(k^2\right)}{k\sqrt{4 R r \sin\theta}} \label{eq:wire} \\
  k = \sqrt{\frac{4 R r \sin\theta} { r^2 + R^2 + 2 r R \sin\theta}} \nonumber
  \label{eq:wire} 
\end{eqnarray}
where $E,K$ are the complete elliptic functions, $R=r_{\rm max}$ is the radius of pressure maximum in the torus,
and $A_0$ is a constant that it is used to scale the magnetic field and the initial $\beta$-parameter across the initial torus.

Another, frequently explored configuration, is that following the constant density contours in the torus, i.e.:
\begin{equation}
  A_{\varphi}={\rho \over \rho_{\rm max}} - \rho_{0}
  \label{eq:loop}
\end{equation}
where we use $\rho_{0}=0.2$.

%unit conversions

Following \cite{janiuk2013}, we use the value of the total initial mass
of the torus to scale the density over the integration space, and
we base our simulations on the physical units. We use
\begin{eqnarray*}
L_{\rm unit}=\frac{G M}{c^2}=1.48\cdot 10^5 \frac{M}{M_\odot} ~~ \rm{cm} \\
T_{\rm unit}=\frac{r_g}{c}=4.9\cdot 10^{-6} \frac{M}{M_\odot} ~~ \rm{s} \\
%f_g= 6.0 \cdot 10^{20}\sqrt{\rho_{torus}} \left(\frac{M}{M_\odot}\right)^2 \rm{Mx}
\end{eqnarray*}
as the spatial and time units, respectively.
Now, the density scale is related to the spatial unit as $D_{\rm unit}=M_{\rm scale}/L_{\rm unit}^{3}$. We conveniently adopt the scaling factor of 
$M_{\rm scale}=1.5\times 10^{-5} M_{\odot}$, so that the total mass contained 
within the simulation volume (mainly in the FM torus) is about 0.1 $M_{\odot}$, for  a black hole mass of 3 $M_{\odot}$. 
%The last scaling is derived by the cgs definition of the magnetic units and the $r_g, t_g, \rho=\rho_{\rm torus} \, \rm{[g\,cm^{-3}]}$ units are assumed.
The $r_{\rm in}$ and $r_{\rm max}$ radii of the torus are
 related to a number of factors that are very difficult to determine in advance.

 %having the fiducial values of $r_{\rm in} = (3.5 - 4.5)\,r_g$ and $r_{\rm max} = (9-11)\,r_g$.
 %(see Table \ref{tab:in}).

\subsection{Time dependent evolution}

Regardless of the specific magnitude of the plasma $\beta-$parameter 
defined as the ratio of the fluid's thermal to the magnetic pressure, $\beta \equiv p_g / p_{\rm mag}$, 
the flow slowly relaxes its initial configuration, becomes geometrically thinner, and launches the outflows.
Also, the quantities as measured on the black hole horizon, namely the mass flux, energy and angular momentum flux, or the magnetic flux, are a function
of time and may exhibit fast variability.

\begin{figure}
    \includegraphics[width=0.9\textwidth]{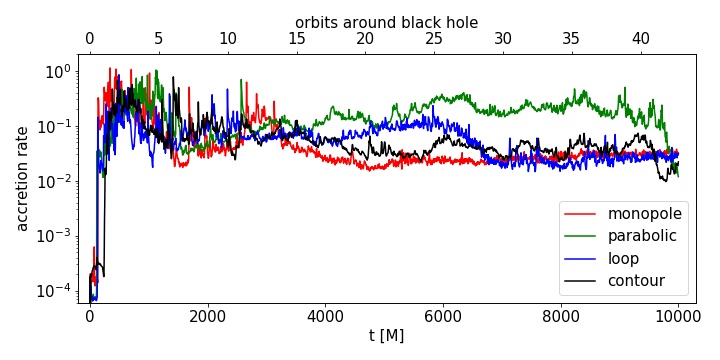}
    \caption{Mass accretion rate onto black hole through the magnetized torus embedded with different initial configuration of the poloidal magnetic field. Dimensionless units are used in the plot: accretion rate through horizon is normalized to unity, and in the time unit $t_{\rm g}=GM/c^{2}=M$ is adopted.}
    \label{fig:polo}
  \end{figure}

Time dependence of accretion rate at the black hole horizon, as resulting from the 2-dimensional simulations of accretion torus evolving under the influence of magnetic field with various initial poloidal configurations, is displayed in Figure \ref{fig:polo}. The configuration of
split monopole were studied in number of works, starting from \cite{bz77}, while the parabolic field in GRMHD simulations of jets in M87 were proposed e.g. by \cite{nakamura18}. The current loop with a simplified formula proposed by \cite{kolos2017} is used for this plot,
and the contour field configuration is that given by Eq. \ref{eq:loop}.

Time variability of the outflows is related to the variability of accretion, and
driven by the same mechanism of magneto-rotational instability
(see \cite{kostas} for details). What we want to measure for the relativistic jet in GRBs, is the quantity related to its energy and the Lorentz factor

%simulation results
The magnetization $\sigma$ and normalized energy, $\mu$, are defined in GR MHD as:
\begin{equation}
 \sigma= \frac{\left(T_{em}\right)^r_t}{\left(T_{m}\right)^r_t} \nonumber
%-\frac{b^\mu b_\mu} 
 \qquad
 \mu = - \frac{T^r_t}{\rho u^r}
\end{equation}
The energy conservation along a field line gives $\mu=\gamma h \left(1+\sigma\right)$, as the sum of the inertial-thermal energy of the plasma, $\gamma h$, and its Poynting flux, $\gamma h \sigma$.
Hence, the maximum achievable Lorentz factor $\Gamma_{\infty} = \mu$, when all the Poynting and the thermal energy is transformed to baryon bulk kinetic ($\sigma \to 0 , \xi \to 1$) (see \cite{vlahakis}).
%(Vlahakis \& Koenigl 2003) [ref].

\begin{figure}
     \includegraphics[width=0.5\textwidth]{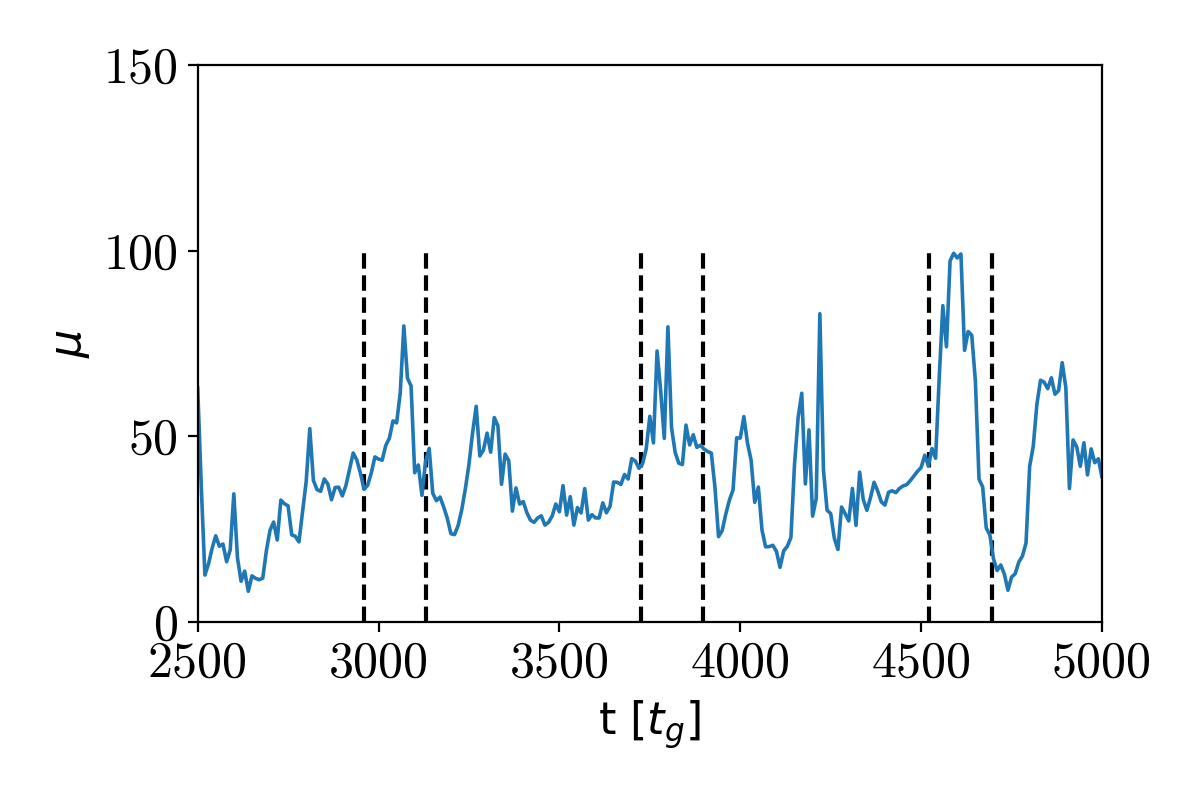}
     \includegraphics[width=0.5\textwidth]{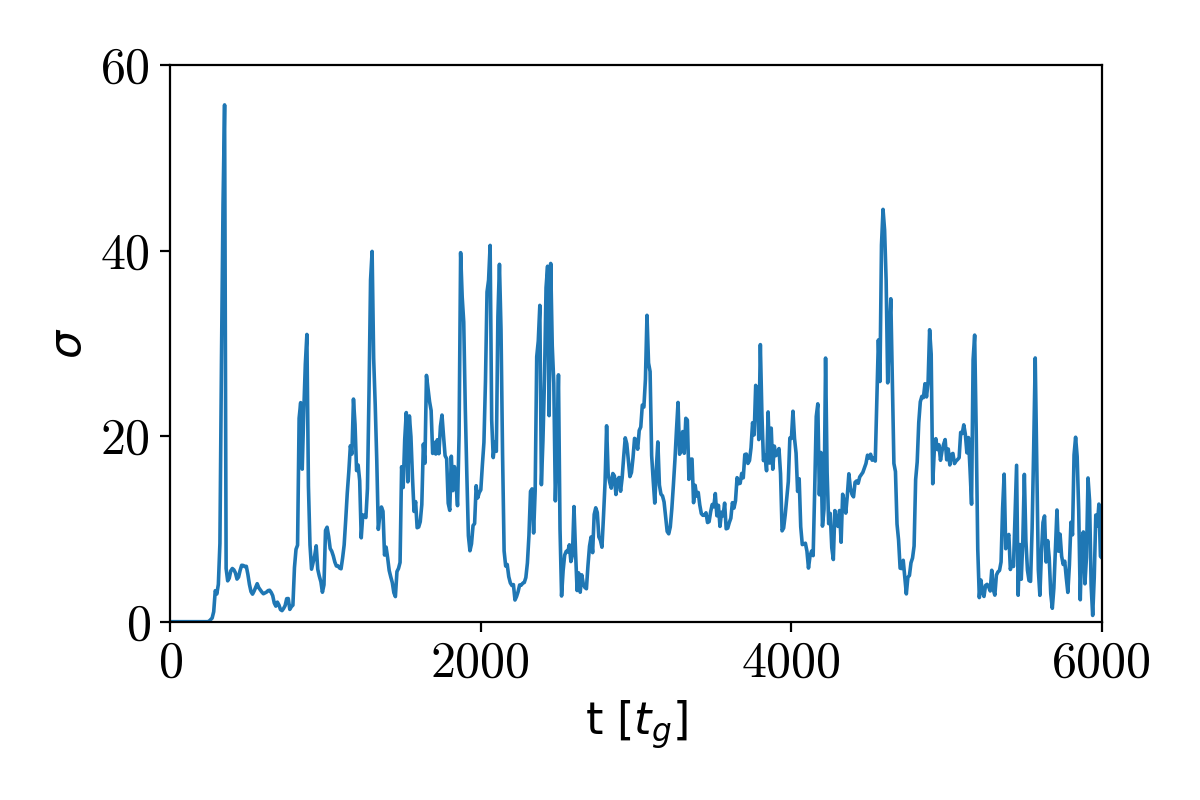}
     \caption{Model MD-Magn from the paper \cite{kostas}.
      {\bf Left:} Time dependence of the energetics parameter $\mu$ (equiv. to $\Gamma_{jet}$ at infinity), measured close to the jet axis. Dashed lines mark the timescale of the MRI instability in the torus. {\bf  Right:} Time dependence of the magnetisation parameter $\sigma$, measured close to the jet axis, over the whole simulation. }
\label{fig:mu_vary}
   \end{figure}

In the numerical code, the variables are distributed on the grid and vary
with time, so variability of $\mu$
can be measured at outer and inner regions of jet.
In Figure \ref{fig:mu_vary} we present an exemplary time sequence of the
jet energetics measured with $\mu$ at specific point of reference:
$p(x,y)=(11.5, 200) r_{\rm g}$. The initial condition for the model was given
by the FM-torus parameterized with the inner radius of $20 r_{g}$ and the radius of pressure maximum of $25 r_{g}$, and located around the Kerr black hole whose spin parameter is $a=0.6$.
The torus is embedded with magnetic field of the initial poloidal configuration as given by formula \ref{eq:wire}, with normalisation parameter $A_{0}=32$ (measured in the units of $f_{\rm g} = 6.0 \cdot 10^{20}\sqrt{\rho_{torus}} \left(\frac{M}{M_\odot}\right)^2 \rm{Mx}$. The last scaling is derived by the cgs definition of the magnetic Maxwellian units, and the $r_g, t_g, \rho=\rho_{\rm torus} \, \rm{[g\,cm^{-3}]}$ units are assumed.

   As the figure shows, the jet is strongly magnetized, which is reflected by the large value of the $\sigma$ parameter. The intense variability of the outflow is correlated with the magneto-rotational instability (see \cite{balbus1991}), and in the exemplary plot one, or two peaks of $\mu$ are present per each MRI interval.
   Here, the timescale $T_{\rm MRI}$ is obtained from the maximum growth
   rate $\Omega_{\rm max}$ for a disk in the Kerr metric (see \cite{Gammie2004})
   and for an observer at infinity it is given by $(9/16) \Omega^{2} (D^{2}/C)$ with $D, C$ coefficients taken from \cite{novikov}.
   The wavelength of the fastest growing mode is given by
   $\lambda_{\rm MRI} = (2\pi /\Omega) v_{A}$, with $v_{\rm A, \theta}=b^\theta / \sqrt(\left(\rho h \right)^2 + b^\mu b_\nu)$ being the $\theta$ component of the Alfv\'{e}n velocity.
   
\subsection{Exemplary results for the NS-NS merger scenario}

The global picture of short GRB progenitor involves the NS-NS merger, after which, and after a possible transition through the hypermassive neutron star phase, the system ends up with configuration composed of a stellar-mass black hole and magnetized torus.
The source of power of the GRB central engine may be
the anihillation of neutrino-antineutrino (e.g. \cite{janiuk2013}), and
the black hole rotation, mediated by magnetic fields.
The rotational velocity of the magnetic field is
$\Omega_F = F_{t\theta}/F_{\theta\phi}$, while the
angular frequency of the black hole $\Omega_{\rm BH}=\left(a / 2\right)\left(1+\sqrt{1-a^2}\right)$. In the innermost region closest to the black hole, the ratio $\Omega_{\rm F}/\Omega_{\rm BH}$ is 0.45-0.55, which is in accordance with the operation of Blandford-Znajek process, \cite{bz77}. 

%jet model results
We present here an exemplary model HS-Magn of \cite{janiuk2019},
  parameterized with the black hole mass of $3 M_{\odot}$ and mass of the torus equal to 0.1 $M_{\odot}$ -- hence the density scaling in cgs units.
As the figure shows, the magnetic fields help launching the uncollimated outflow from the surface of accretion torus.

   \begin{figure}
     \includegraphics[width=0.5\textwidth]{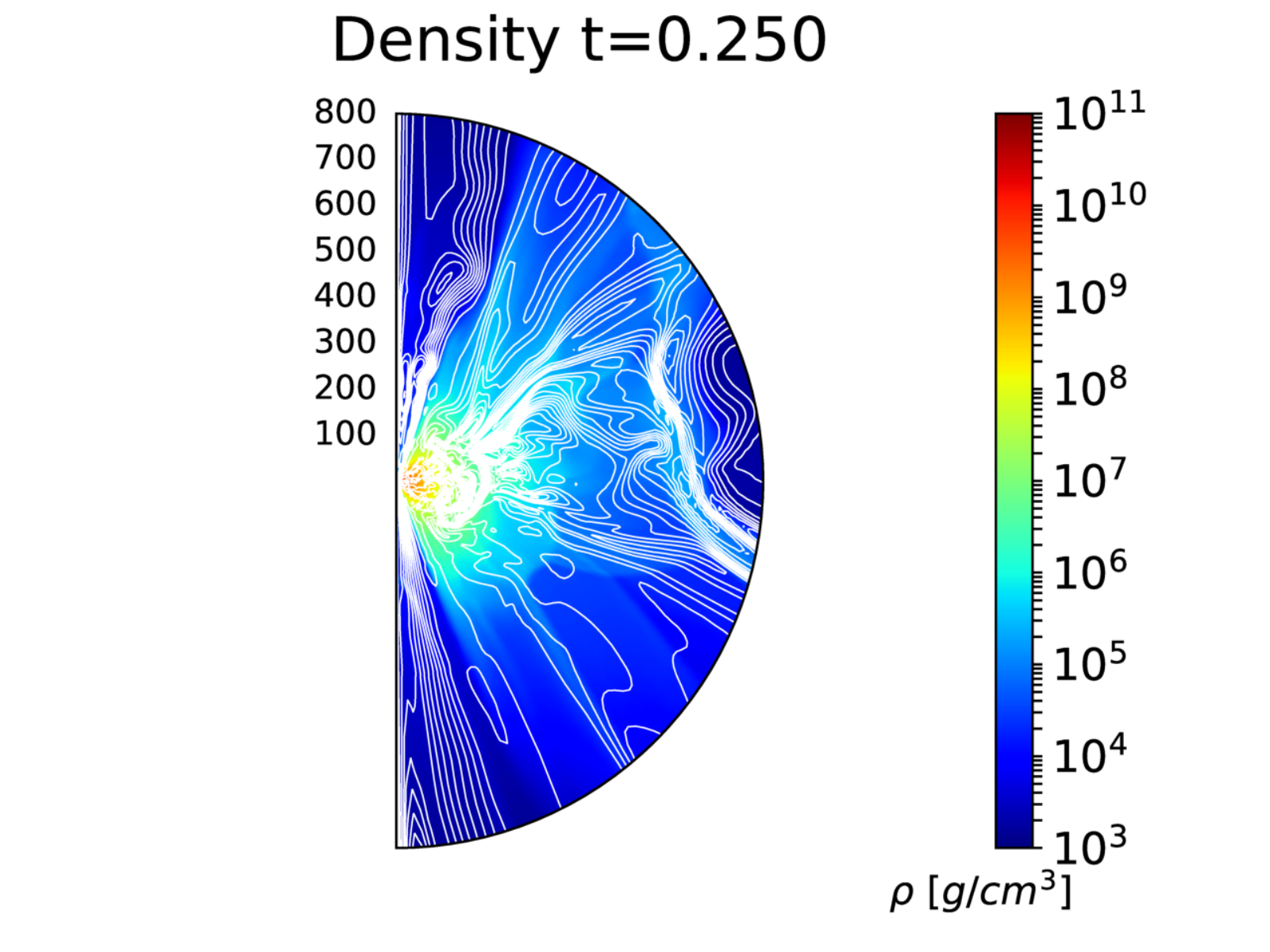}
     \includegraphics[width=0.5\textwidth]{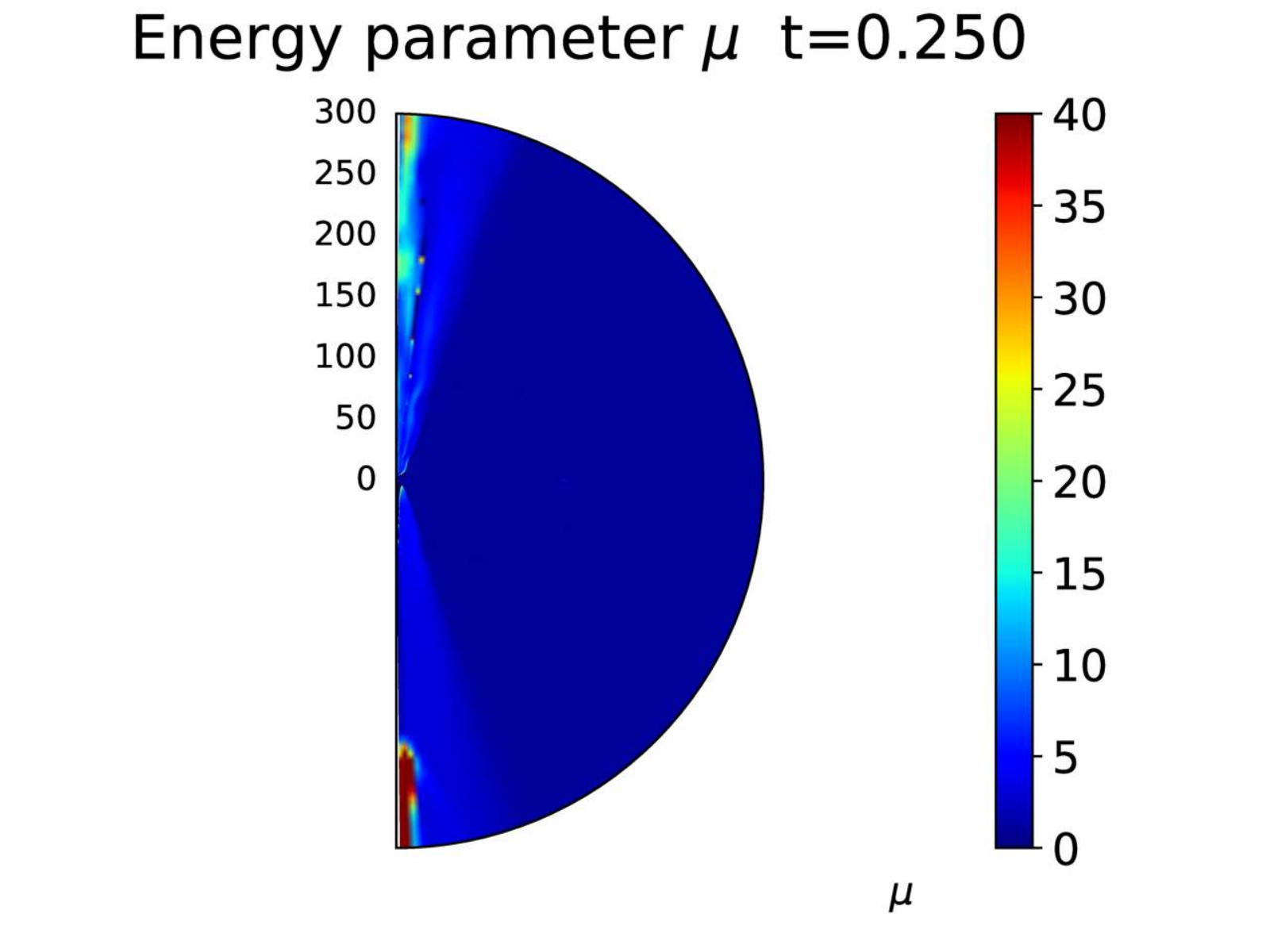}
     \caption{Distribution of density and magnetic field at simulation
       time 17000 $t_{\rm g}$, which corresponds to 0.25 s after the BH formation.
       The parameters are 3 Solar mass black hole with a spin of $a=0.9$. The initial gas to magnetic pressure ratio within the FM torus was assumed $\beta_{\rm init}=10$, and the initial poloidal
       field had a shape of density isocontours. Mass of the torus was 0.1 $M_{\odot}$. Left plot shows the contours of constant vector potential $A_{\phi}$, and the right panel shows the distribution of jet energetics parameter $\mu$. Spatial scale of the plot is given in gravitational radii, $r_{\rm g}$.}
\label{fig:GRB_lc}

   \end{figure}
   The model is parameterized with black hole spin $a=0.9$, and
   initial $\beta=10$ parameter, defined as the ratio of the fluid's
   thermal to the magnetic pressure, $\beta \equiv p_{\rm g} / p_{\rm mag}$ for the torus configuration.
%while every set includes models differing with the black hole spin.
  
   \section{Nucleosynthesis in the outflows from GRB engine}
   
   In the GRB central engine, when we change from dimensionless, to cgs units, we
   deal with the hyperaccretion rates of 0.01-10 $M_{\odot}/s$.
   The steady state and time-dependent models were proposed from 1990's
   \cite{popham1999, dimatteo2002, Kohri2005, ChenBeloborodov2007, Janiuk2007, JaniukYuan2010}, showing that the equation of state (EOS) of such gas is not ideal, but the plasma is composed of free $n$, $p$, $e^{+}$, $e^{-}$ particles and that the chemical and pressure balance is followed from nuclear reactions:
electron-positron capture on nucleons, and neutron decay \cite{reddy1998}.
%(Reddy, Prakash \& Lattimer 1998)
Adopting the charge neutrality condition, and adjusting for the mass fraction of free nuclei with respect to the Helium nuclei, that also form,
we can compute the number  densities of free species. Their degeneracy parameters are determined also for a given temperature and density in the torus, and the gas obeys the Fermi-Dirac statistics.
The torus is cooled by neutrinos, and due to the
neutrino absorption and scattering, the neutrinos can become partially trapped in the torus. These neutrino energies peak in MeV range \cite{wei2019}
%(Wei et al. 2019)
so they are not detectable by the current neutrino detectors on Earth. Nevertheless, the anihillation of neutrinos  and antineutrinos over the jet axis can give an extra source of power to the relativistic jets in GRBs (and both are larger for a higher black hole spin).

%{Di Matteo et al. 2002; Kohri et al. 2002, 2005; Chen \& Beloborodov 2007;
%Janiuk et al. 2007; Janiuk \& Yuan 2010; Lei et al. 2009; Janiuk et al. 2013; Liang et al. 2015; Janiuk 2017, 2019
The code HARM-COOL is computing the evolution of the GRB central engine as driven by the GR MHD equations with the methods described above. This code version has been developed in our group, and is supplemented with the module for the EOS of dense and degenerate matter, as adequate for physical conditions in the
short GRB engine. The module works
thanks to a non-trivial transformation between conserved variables and
primitive variables in HARM \cite{Noble2006}.
The EOS is being computed dynamically during the simulation, and stored in
tables, over the pre-defined large range
of density and temperature. The \textit{pthread} interpolation with the spline method is used to determine pressure and internal energy from the tabulated EOS.

We find that the matter is neutronized, and the electron fraction
$Y_{\rm e}=n_{p}/(n_{p}+n_{n}) < 0.5$.

In the torus, the nuclear statistical equilibrium conditions allow for a
synthesis of further heavy isotopes, up to mass number $A\sim 80$, i.e.,
the most abundant is the Iron group, with $^{54}$Fe, $^{55}$Co, and $^{56}$Ni, while some heavier isotopes like $^{74}Se$, $^{80}Kr$, and $^{84}Sr$, and $^{90}Zr$ are also present with the relative abundances of about $10^{−12}-10^{−10}$
\cite{janiuk2014}.
%These results are based on density, temperature and electron fraction
%profiles from 1-D model and postprocessed with the code ??? by B. Meyer (1994)

\begin{figure}
  \includegraphics[width=0.8\textwidth]{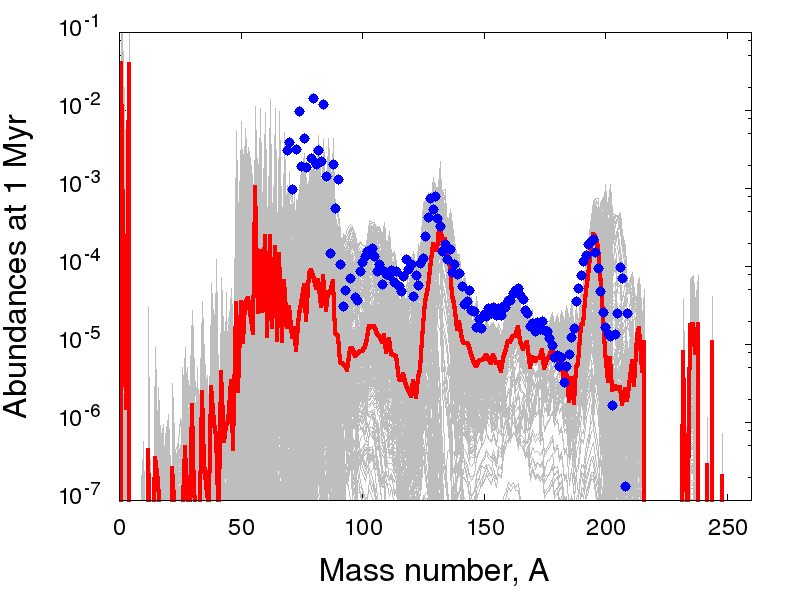}
  \caption{The heavy element relative abundance pattern, calculated at the tracer particles on the outflow from the accretion disk around Kerr black hole (spin $a=0.9$). Gray lines represent individual outflow tracers, and red line is the angle-averaged profile. Blue points show the Solar abundance data.}
  \label{fig:abund}
  \end{figure}

To reproduce the elements abundance pattern as synthesized in the outflow from
GRB central engine, we need to relax the statistical equilibrium condition.
The r-process elements are formed in dynamical outflows via capture of free neutrons, when the timescale of this process is much shorter than the
equilibrium timescale.
Nucleosynthesis in dynamical outflows is performed in our simulation via post-processing of the results\footnote{We use the publicly available code SkyNET
\cite{lippuner}}. We store the data for the outflow density, temperature, and electron fraction, on the outflow trajectories. They are defined by a dense set of tracer particles, which are identified as those leaving the computational domain through the outer boundary.
Some of these particles carry the electron fraction as low as $Y_{\rm e}=0.1-0.2$, and their velocity can be in the range $v=0.2-0.3 c$.
These thermodynamic properties allow for the formation of the r-process elements until the mass number of $A\sim 200$, and reproducing the second and third peak of the Solar abundance pattern (see Figure \ref{fig:abund}). 

According to \cite{cowper2017}, the kilonova UV, optical, and NIR Light curves of GW-GRB 170817 are well modeled by the two-component model for r-process heating. The dynamical ejecta from compact binary mergers, $M_{\rm ej}\sim 0.01 M_{\odot}$, can emit about $10^{40}-10^{41}$ erg/s in a timescale of 1 week, while the subsequent accretion can provide bluer emission, if it is not absorbed by precedent ejecta \cite{tanaka2016}. %(Tanaka M., 2016)
%{Cowperthwaite et al., 2017}
Our modeling is in great agreement with these predictions.
%(Lippuner \& Roberts 2017), used for the nuclear reaction networks

 \section{Summary}

 The potential electromagnetic counterparts of compact object binary mergers are a function of the observer viewing
angle. Rapid accretion of a centrifugally supported disk powers a collimated relativistic
jet, which produces a short GRB.
The working theory of the short GRB signals
in multimessenger era has to cover both the Central Engine and jets physics.
In Central Engine, both accretion and wind ejection are playing role
(MRI turbulence, magnetically/neutrino driven winds)
The observables are: emission from jets (probed by their energetics, minimum time variability scales), and the emission from afterglows, including now the kilonovae. The latter may bring information on accretion physics.
The gravitational waves give new window and relate progenitor properties with the GRBs prompt phase (i.e. can help constrain the mass and spin of the black hole in the engine).
 
In our work, the Central Engine and outflows in short GRB
were studied with respect to their unique properties, with the newly developed version of our Code HARM-COOL. The code
is now available for the community to download and use. By means of this tool, the user may study unique properties of GRBs.
  Recently the code has been used in a number of applications, published in the specialist literature.

{\bf Acknowledgements}
We acknowledge partial support from grant DEC-2016/23/B/ST9/03114
awarded by the Polish National Science Center.
The Interdisciplinary Center for Mathematical Modeling is acknowledged for the computational time on Okeanos supercomputer through grants GB70-4 and GB79-9.

%\end{document}

\bigskip
\bigskip
\noindent {\bf DISCUSSION}

\bigskip
\noindent {\bf JAMES BEALL:} Can you comment on how much computational power will be needed to run a full 3-D simulation with your code?

\bigskip
\noindent {\bf AGNIESZKA JANIUK:} The HARM-COOL code can be run in 2D or 3D setup, and with or without the neutrino-cooling and nuclear EOS module.
We have run the 2-D simulations with the nuclear EOS (grid resolution 256x256)
until time 20,000 dynamical times, and the simulation used 64x24 CPU on the Okeanos supercomputer in Warsaw, for about 2 weeks per each model. The code has to be restarted every 48 hrs due to the cluster occupancy and queueing system. Such long computation is essential to find enough number of tracer
particles that are in the outflow (i.e., these which
leave the GRB engine and are sites for nucleosynthesis).
As for the 3-dimensional simulations, we have ran them only with the adiabatic
EOS, but with resolution 288x256x256, on the cluster 48x48 CPU,
until time of 10,000 dynamical times for 6 days. Including EOS calculations and tracer outflows in the 3D setup is beyond the capability of the Warsaw ICM
supercomputing power.

\bigskip

\end{document}